\documentclass{emulateapj}
\usepackage{apjfonts}
\usepackage{epsfig}

\shorttitle{A Radio-loud Magnetar in X-ray Quiescence}
\shortauthors{Levin et al.}

\begin{document}

\title{A Radio-loud Magnetar in X-ray Quiescence}

\author{Lina Levin\altaffilmark{1,2}, Matthew Bailes\altaffilmark{1}, Samuel Bates\altaffilmark{3}, N. D. Ramesh Bhat\altaffilmark{1},\\
 Marta Burgay\altaffilmark{4}, Sarah Burke-Spolaor\altaffilmark{1,2}, Nichi D'Amico\altaffilmark{4,6}, Simon Johnston\altaffilmark{2},\\
 Michael Keith\altaffilmark{2}, Michael Kramer\altaffilmark{3,5}, Sabrina Milia\altaffilmark{4,6}, Andrea Possenti\altaffilmark{4},\\
  Nanda Rea\altaffilmark{7}, Ben Stappers\altaffilmark{3}, Willem van Straten\altaffilmark{1}}

\altaffiltext{1}{Centre for Astrophysics and Supercomputing, Swinburne University of Technology, Mail H39, PO Box 218, VIC 3122, Australia}
\altaffiltext{2}{Australia Telescope National Facility - CSIRO, PO Box 76, Epping, NSW 1710, Australia}
\altaffiltext{3}{Jodrell Bank Centre for Astrophysics, University of Manchester, Alan Turing Building, Oxford Road, Manchester M13 9PL, United Kingdom}
\altaffiltext{4}{INAF/Osservatorio Astronomico di Cagliari, localit\'a Poggio dei Pini, strada 54, 09012 Capoterra, Italy}
\altaffiltext{5}{Max Planck Institut f\"{u}r Radioastronomie, Auf dem H\"{u}gel 69, 53121 Bonn, Germany}
\altaffiltext{6}{Dipartimento di Fisica, Universit\'a degli Studi di Cagliari, Cittadella Universitaria, 09042 Monserrato (CA), Italy}
\altaffiltext{7}{Institut de Ci\'encies de l'Espai (CSIC-IEEC), Campus UAB, Facultat de Ci\'encies, Torre C5-parell, 08193 Barcelona, Spain}

\begin{abstract}
As part of a survey for radio pulsars with the Parkes 64-m
  telescope we have discovered PSR\,J1622--4950, a pulsar
  with a 4.3-s rotation period. Follow-up observations show that the
  pulsar has the highest inferred surface magnetic field of
  the known radio pulsars (B$\sim$3$\times$10$^{\bf14}$\,G),
  exhibits significant timing noise and appears to have an inverted spectrum.
  Unlike the vast majority of the known pulsar population, PSR\,J1622--4950
  appears to switch off for many hundreds of days and even in its on-state
  exhibits extreme variability in its flux density. Furthermore,
  the integrated pulse profile changes shape with epoch.
  All of these properties are remarkably similar to the only
  two magnetars previously known to emit radio pulsations. 
  The position of PSR\,J1622--4950 is coincident with an X-ray source that,
  unlike the other radio pulsating magnetars, was found to be in quiescence. 
  We conclude that our newly discovered pulsar is a magnetar - the first 
  to be discovered via its radio emission.

\end{abstract}


\section{Introduction}

Magnetars are slowly rotating neutron stars 
that, in contrast to ordinary pulsars, are not powered by their
spin-down energy losses, but 
by the energy stored in their
extremely large magnetic fields, typically $\gtrsim$ 10$^{14}$\,G \citep{Dun92}. 
They are divided into two groups,
Soft Gamma-ray Repeaters (SGRs) and Anomalous X-Ray Pulsars (AXPs),
whose relation to each other and 
to other kinds of neutron stars is still a matter of debate \citep{Mer08}.
SGRs were discovered as sources emitting 
short repeating bursts in the hard X-ray/soft $\gamma$-ray band, 
while AXPs were first detected in the soft X-ray band. Recently, sources have 
been found that show properties of both groups, suggesting that 
AXPs and SGRs belong to the same family (\citealp{Rea09}; \citealp{Mer09}). 
Many attempts have been made to find pulsed radio emission from
magnetars (see e.g. \citealp{Kri85}; \citealp{Gae01}; \citealp{Bur06}) but it was not until 2006 that the first detection
was reported \citep{CamNature06}. To date only two AXPs have confirmed
radio pulsations,
namely XTE\,J1810--197 \citep{CamNature06} and 1E\,1547--5408 \citep{CamApJ666}.
Both of these sources are transient AXPs, a subgroup of magnetars that
occasionally undergo very bright X-ray outbursts. 
Their pulsed radio emission 
was discovered in connection with these outbursts,
and is observed to fade with time after an outburst. 

Here we report on the discovery of PSR J1622--4950, the first magnetar 
discovered in the radio band, and discuss its properties in relation to the two
magnetars previously known to emit radio pulsations.

\section{Observations and Historical Data}
\subsection{Parkes Observations}
The High Time Resolution Universe (HTRU) survey for pulsars and fast
transients is currently being carried out at the Parkes 64-m radio
telescope
\citep{Kei10},
using the 20-cm multibeam receiver \citep{Sta96}.
In brief, an effective bandwidth of 341\,MHz, centered around 1.35\,GHz, 
is divided into 874 frequency channels and sampled using 2 bits every 64\,$\mu$s.
In April 2009 (MJD = 54939) a bright radio pulsar, PSR\,J1622--4950, was discovered.
It has a period ($P$) of 4.326\,s and a dispersion measure (DM) of 
820\,cm$^{-3}$\,pc, corresponding to a distance of $\sim$ 9\,kpc, 
if the Cordes--Lazio model for the distribution 
of the electrons in the interstellar medium is assumed \citep{Cor02}.

\begin{figure*}[]
	\begin{center}
		\includegraphics*[width=12cm]{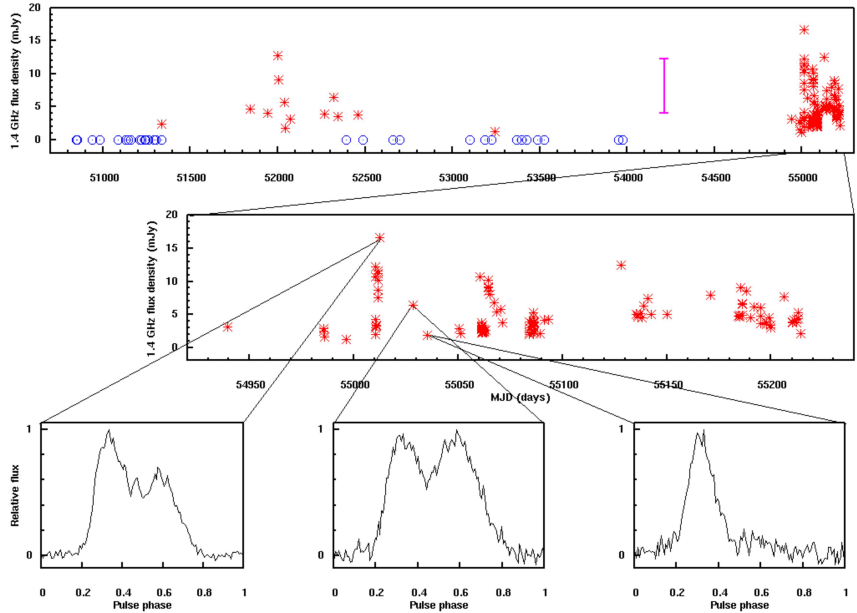}
	\end{center}
	\caption{Light curve and profile variations for PSR\,J1622--4950. The top plot shows the temporal variation of flux density at 1.4\,GHz. Archival timing data collected within the framework of the Parkes Multibeam Survey ends just before MJD = 54000, and the discovery of PSR\,J1622--4950 by the HTRU pulsar survey was made on MJD = 54939. The red asterisks show the detections (with errors smaller than the size of the points), and the blue circles indicate observations during which PSR\,J1622--4950 was not detected (down to a limiting flux density of 1.2\,mJy). The purple error bar indicates the observation centered at 6.3\,GHz from the Methanol Multibeam survey that detected PSR\,J1622--4950. When converted into a 1.4 GHz flux density, the large error is dominated by the uncertainty in the spectral index; the top value of the bar corresponds to a flat radio spectrum and the bottom value is derived from our best estimate of the spectral index. The bottom plots show three different pulse profiles from three consecutive observations, taken on 30 June 2009, 16 July 2009 and 23 July 2009.\\}
	\label{Fig:Lightcurve}
\end{figure*}

After the HTRU discovery we embarked on a timing
campaign, observing PSR\,J1622--4950 a total of 110 times at 1.4\,GHz
and 10 times at 3.1\,GHz, obtaining polarimetric information on most occasions.
The digital filterbank system (DFB3) used to create the folded profiles first converts the 
analogue voltages from each polarization channel of the linear
feeds into digital signals. It then produces 1024 polyphase filterbank frequency channels
that are folded at the apparent topocentric
period of the pulsar into 1024 pulsar phase bins, and written to disk every 20\,s.
Four Stokes parameters are recorded.
To determine the relative gain of the two
polarization channels and the phase between them, a calibration signal 
is injected at an angle of 45$^\circ$ to the feed probes. 
The data are analyzed off-line using the {\sc psrchive} package\footnote{See http://psrchive.sourceforge.net}
\citep{Hot04}
and corrected for parallactic angle and the orientation of the feed. 
The position angles are also corrected for Faraday rotation through the interstellar
medium using the nominal rotation measure.

\subsection{Archive Mining}
The area of sky containing PSR\,J1622--4950 was 
covered previously by the Parkes Multibeam Pulsar Survey \citep{Man01}
but, somewhat surprisingly given its large apparent flux density and DM, 
the pulsar was not detected in these data.
However, that survey did find two other radio pulsars in 
close proximity to PSR\,J1622--4950 and monitored them for a number of years,
namely PSRs J1623--4949 and J1622--4944 at an angular separation 
of 11$'$ and 7$'$ respectively.
These archival data were reprocessed using the period and DM of PSR\,J1622--4950. 
The first detection was made in observations recorded in June 1999 (MJD = 51334), and it was then detected in 11 
observations between October 2000 (MJD = 51844) and July 2002 (MJD = 52458) (see the top panel of Fig. \ref{Fig:Lightcurve}).
However, in the 14 subsequent observations preceding August 2006 (MJD = 53975), 
the pulsar was detected only once. No further archival data of the two nearby pulsars were available after that date.
The Methanol Multibeam pulsar survey at 6.3\,GHz 
\citep{Bat10}
also covered the
part of the sky containing PSR\,J1622--4950, and
by reprocessing the relevant data, the pulsar was detected in 
an observation taken in April 2007 (MJD = 54211). 

\subsection{ATCA and Chandra Observations}
Observations of the pulsar were made on 8 Dec 2009 and 27 Feb 2010 with 
the Australia Telescope Compact Array (ATCA), an east-west synthesis telescope 
located near Narrabri, NSW, which consists of six 22-m antennas on a 6-km track. 
The observations were carried out simultaneously at 5.5 and 9.0 GHz with 
a bandwidth of 2 GHz at each frequency subdivided into 2048 spectral channels, 
and full Stokes parameters. 
The source was tracked for 12-hours in each of the EW352 and 750B array configurations.

Initial data reduction and analysis were carried out with the
MIRIAD package\footnote{See http://www.atnf.csiro.au/computing/software/miriad} using standard techniques. 
After flagging bad data, the primary calibrator (PKS 1934-638) was used for flux density
and bandpass calibration and the secondary calibrator (PKS J1613-586)  
was used to solve for antenna gains, phases and polarization leakage terms.

\begin{figure}[]
	\begin{center}
		\includegraphics*[width=7cm]{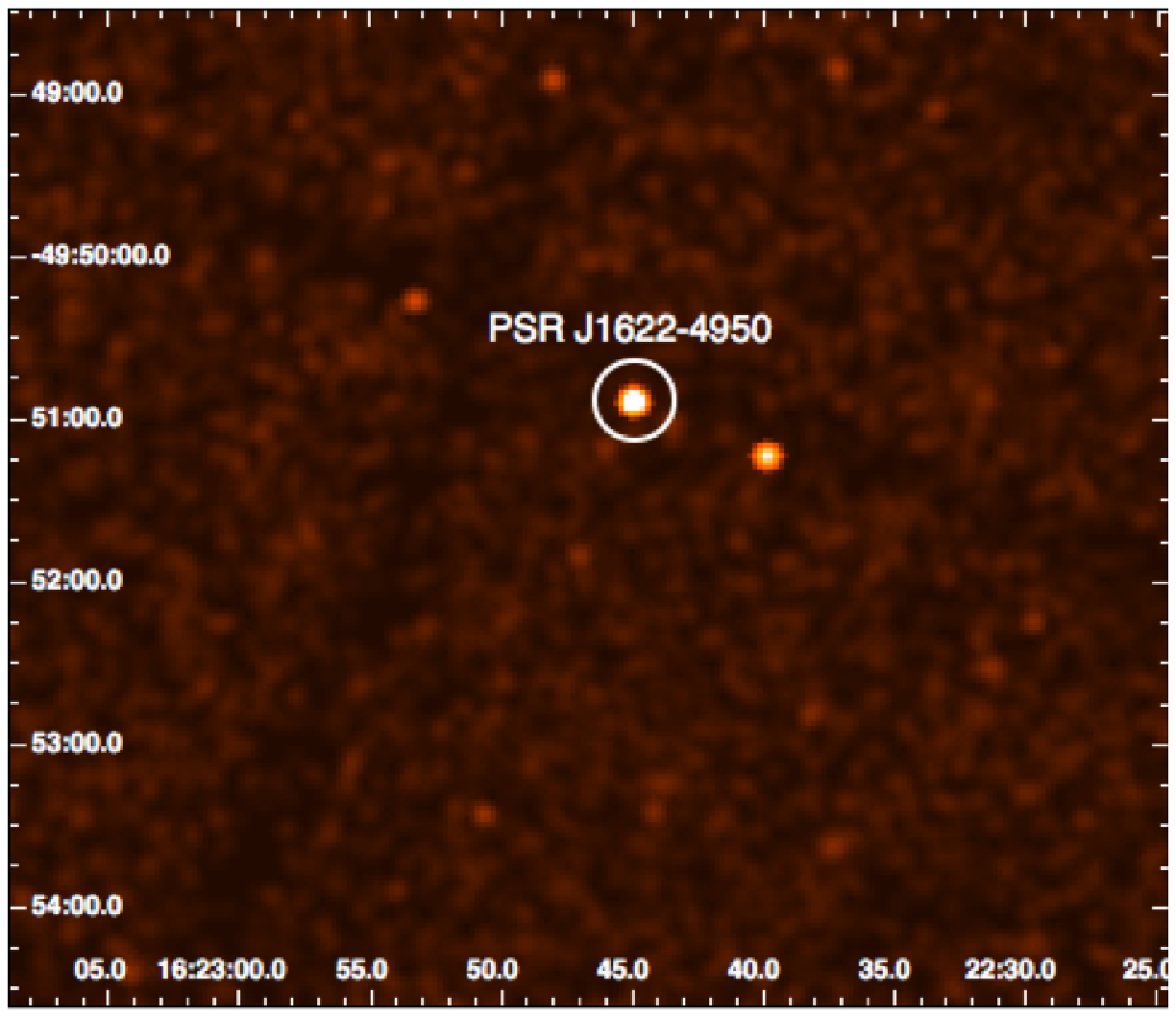}
		\includegraphics*[width=7cm]{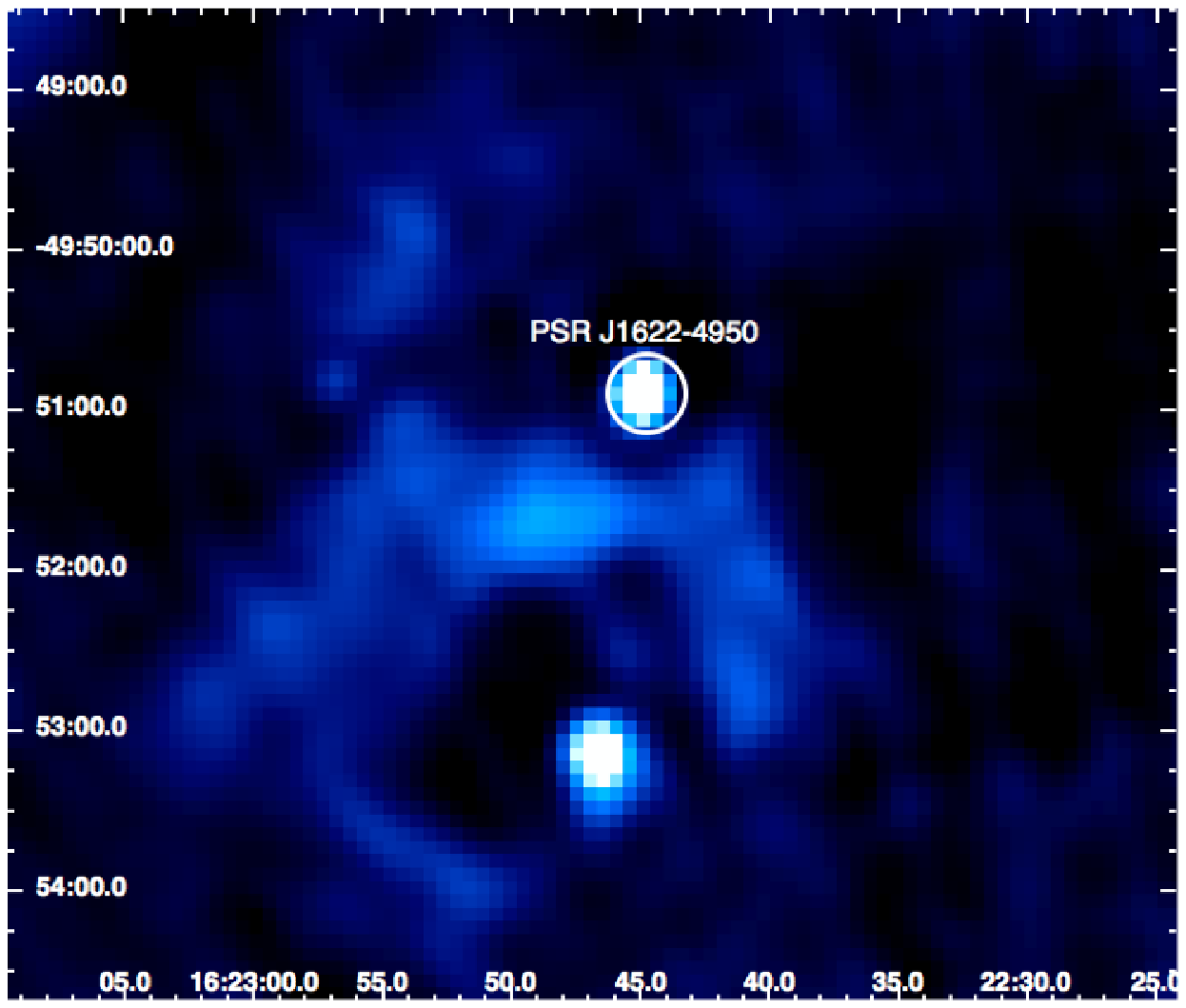}
	\end{center}
	\caption{
	{\it Top:} {\it Chandra} X-ray observations of the region around PSR J1622--4950. The source encircled with a 
	15$''$ marker, CXOU J162244.8--495054, is coincident with the radio point source 
	in the ATCA image.
	{\it Bottom:} ATCA radio observations at 5.5 GHz. The encircled source is PSR J1622--4950. 
	South of the magnetar is a faint ring of 
	extended emission that appears to be non-thermal and may be a supernova remnant (see text for details).\\}
	\label{Fig:ATCA-CXO}
\end{figure}

We also obtained a 20-ks {\it Chandra} observation performed
with ACIS-I on 10 July 2009 to search for an X-ray counterpart.
Figure \ref{Fig:ATCA-CXO} shows the resultant radio and X-ray images. Although there
are a number of radio and X-ray sources present in the field of view,
a highly polarized radio point source is coincident with the brightest 
X-ray source, CXOU J162244.8--495054,
which is also the counterpart of a source seen in archival ASCA data, AX J162246--4946.
We note that CXOU J162244.8--495054 was one of only two X-ray sources 
in the field without an optical/infrared counterpart. We are therefore
confident that the polarized source seen in the radio is indeed
the pulsar and that it has an X-ray counterpart.

\section{Analysis, Results and Discussion}

\subsection{Radio Light Curve}
Throughout the HTRU timing campaign,
the pulsar was detected in every observation with a greatly 
varying flux density, which fluctuated by as much as a factor of $\sim$6 within a 24--hour period.
These flux variations cannot be attributed to
interstellar scintillation and must be intrinsic to the source, as at 20cm
wavelengths, pulsar flux densities are reasonably stable at such large
DMs \citep{Sti00}.
The pulse profile averaged over each observation (typically 5-10 minutes)
also changes
shape on short timescales, often from day to day. 
Although such variability in both the integrated profile and
the flux density is very uncommon in pulsars,
it is a distinctive feature of the two magnetars whose pulsed 
emission has been detected in the radio band (\citealp{CamApJ666}; \citealp{Ser07}; \citealp{Kra07}).
The light curve for PSR\,J1622--4950 and three representative pulse 
profiles are shown in Fig. \ref{Fig:Lightcurve}.

\subsection{Timing}
To attain a stable timing solution of a source
that is changing its pulse profile from observation to observation,
special care needs to be taken in the timing procedure. 
In conventional pulsar timing a standard profile is created and used to
calculate each pulse's time of arrival (TOA) at the telescope.
In the case of PSR\,J1622--4950 each pulse profile is different, but we noticed that all profiles 
seem to consist of the same components, only differing by their amplitudes
(and sometimes the amplitude for a specific component was indistinguishable from
zero). Software from the {\sc psrchive} package \citep{Hot04}
was used to fit gaussians to one of the profiles that showed all the different components and a
model was made to describe how the components build up the profile.
This model was then used to calculate TOAs, by letting the amplitudes of the different
components vary.

When TOAs for all our observations were calculated, the {\sc Tempo}
software\footnote{See http://www.atnf.csiro.au/research/pulsar/tempo/} 
was used to analyze the rotational history of PSR\,J1622--4950. 
The resultant fit requires at least 9 period derivatives to obtain reasonable
phase connection for all our data, which is not
uncommon for magnetars \citep{CamApJ679}. 
To monitor variations of the period derivative with time, we have also measured 
 ${P}$ and $\dot{P}$ over several time intervals separately. 
The time spans were chosen to be as long as possible 
while keeping phase connection without adding higher period derivatives. 
This usually resulted in groups of TOAs spanning about 40 days each. 
It appears that $\dot{P}$ is fluctuating within a factor of $\sim$2 in time. 
From this analysis we derived an average period derivative, 
$\dot{ P}$ = 1.7$\times$10$^{-11}$\,s\,s$^{-1}$. 
The parameters from the timing analysis are listed in Table \ref{Tab:Summary}.

\begin{figure}[]
	\begin{center}
		\includegraphics*[width=8.5cm]{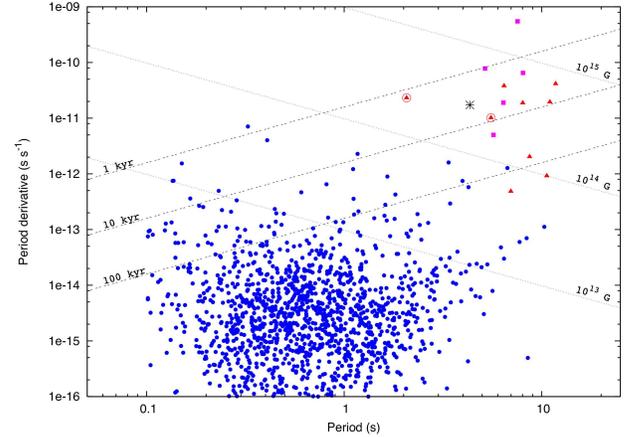}
	\end{center}
	\caption{Extract of the $P$-$\dot{P}$ diagram, showing the longer period pulsars and those with higher inferred magnetic field strengths. Blue dots represent ordinary pulsars, red triangles are AXPs, purple squares are SGRs, and the black asterisk represents PSR\,J1622--4950. The two encircled AXPs are the magnetars that are observed to emit radio pulsations. Dashed lines correspond to lines of constant characteristic age and dotted lines to constant magnetic field strength. PSR\,J1622--4950 has a slightly higher magnetic field strength than the AXPs that are pulsating in the radio band, but clearly higher than that of any of the ordinary pulsars.\\}
	\label{Fig:PPdot}
\end{figure}

The timing solution for PSR\,J1622--4950
implies a very high surface magnetic field strength of
${B}$ $\equiv$ 3.2$\times$10$^{19}$\,G
$\sqrt{ P \dot{ P}}$ $\approx$ 2.8 $\times$ 10$^{14}$\,G
and a characteristic age of
$\tau_{\rm c} \equiv  P/(2\dot{P}) \approx$ 4000\,years \citep[e.g.][]{Lor05}.
This is the highest surface magnetic field of any radio pulsar known to date. 
Figure \ref{Fig:PPdot} shows an extract of the $P$-$\dot{P}$ diagram, where
the currently known magnetars are clustered in the upper right corner, with 
$P$ $\gtrsim$ 2\,s and $ \dot{P} \gtrsim$ 5$\times$10$^{-13}$\,s\,s$^{-1}$.
The two radio emitting magnetars are found in the lower parts of 
the magnetar group, with periods that are short in comparison to the 
other AXPs and inferred magnetic field strengths in the lower half of 
the magnetar range, adjacent to the ordinary radio pulsars with 
the highest surface magnetic fields. The gap between these two groups 
of pulsars is narrowing with new discoveries, such as the detection of magnetar-like X-ray bursts
from PSR\,J1846--0258 \citep{Gav08}, which was previously thought to be solely rotation-powered.
PSR\,J1622--4950 is found in the same region of the plot as the two radio emitting magnetars.

\subsection{Polarization}
\begin{figure}[]
	\begin{center}
		\includegraphics*[width=8cm]{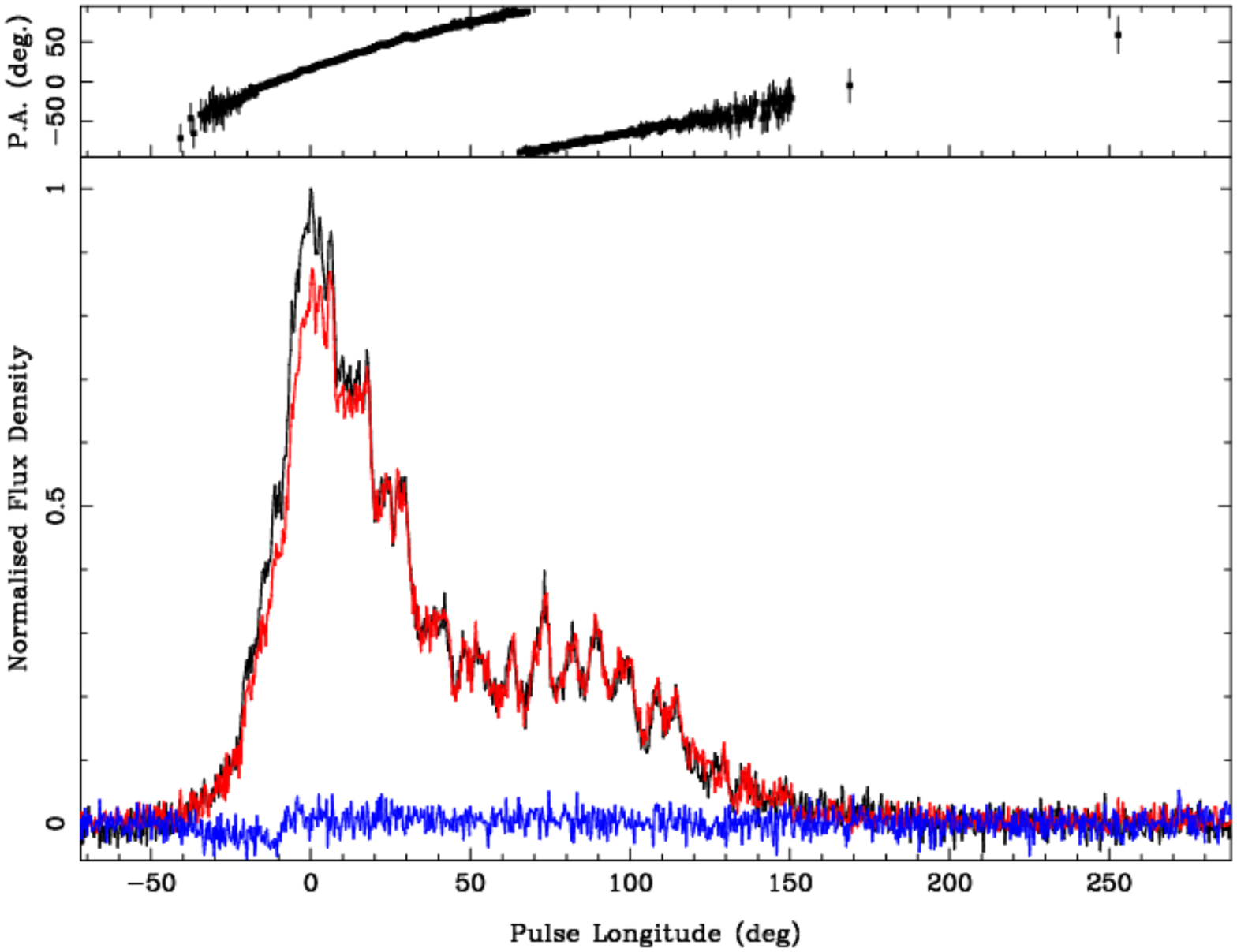}
		\includegraphics*[width=8cm]{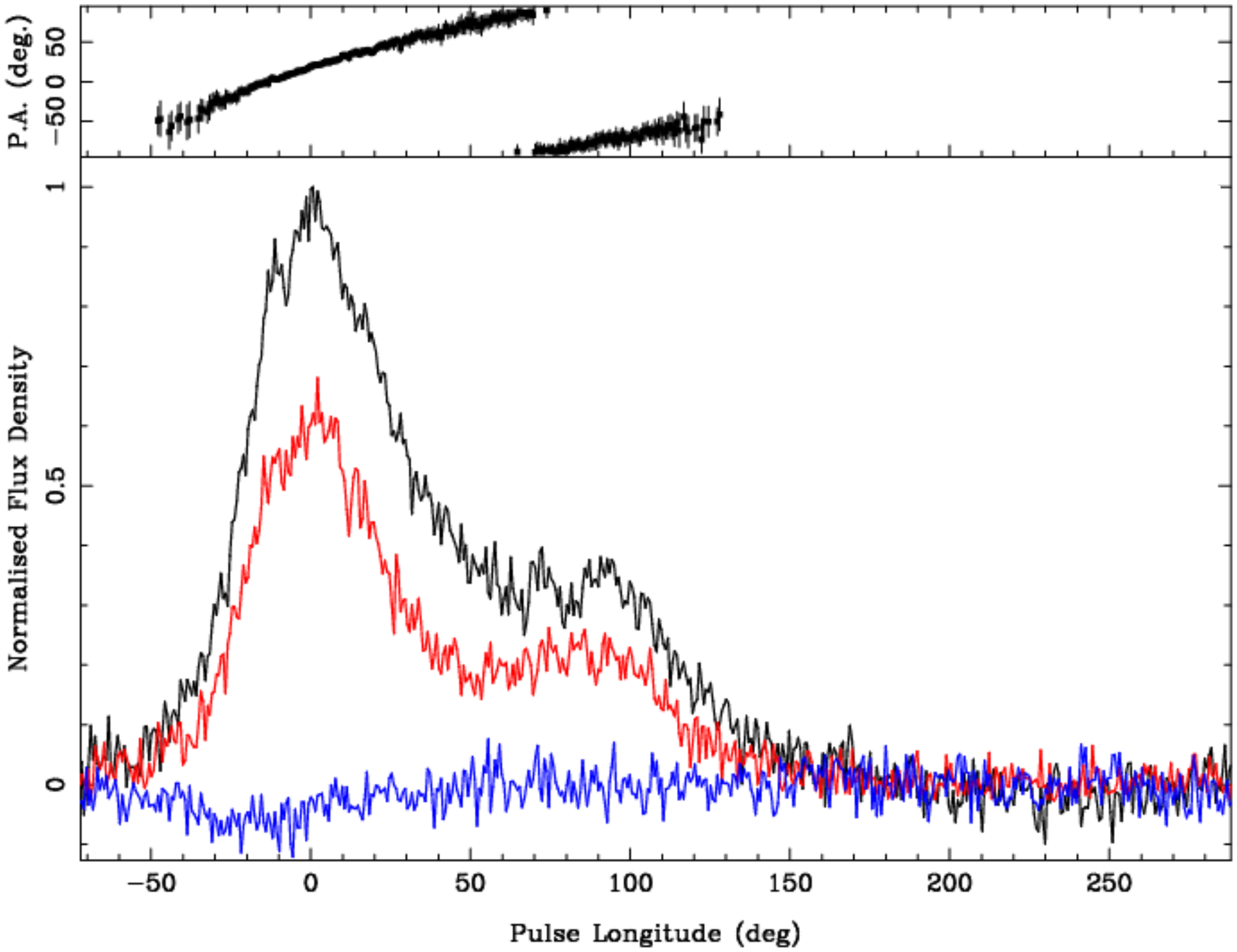}
	\end{center}	
	\caption{Polarimetric profiles of PSR\,J1622--4950 at 3.1\,GHz (top) and 1.4\,GHz (bottom) observed at Parkes on the same day (MJD = 55197). The blue and red lines represent circular and linear polarization respectively, while the black lines represent the total intensity. The full 360$^\circ$ of pulse rotation are shown, with the zero point of the pulse longitude set at the peak value of the flux density. Absolute position angles at infinite frequency are shown after correction for a rotation measure of --1484\,rad\,m$^{-2}$.\\}
	\label{Fig:Polarisation}
\end {figure}

The emission from the magnetar is 
often highly linearly polarized, typically with greater polarization 
at 3.1\,GHz than at 1.4\,GHz (see Fig. \ref{Fig:Polarisation}).  
The pulsar also intermittently shows 
significant circular polarization.
The Faraday rotation measure derived from these observations is very 
large, RM\,=\,--1484$\pm$1\,rad\,m$^{-2}$.
A detailed analysis of these polarization data will be reported elsewhere but we note
the similarity between the polarization properties of PSR\,J1622--4950 and the other radio pulsating magnetars 
1E\,1547--5408 \citep{CamApJ679} and XTE\,J1810--197 (\citealp{Kra07}; \citealp{CamApJ659}).

\subsection{Radio Spectrum}
We combined the various flux density measurements to estimate the spectral
index of the pulsar in the radio band. The average flux density
at 1.4\,GHz is $\langle$$ S_{1400}$$\rangle$ = 4.8\,mJy 
(with standard deviation $\sigma_{ S_{1400}}$ = 2.8\,mJy) and at
3.1\,GHz is $\langle$$ S_{3100}$$\rangle$ = 4.9\,mJy ($\sigma_{ S_{3100}}$ = 2.5 mJy).
The ATCA measurements yield $\langle$$ S_{5500}$$\rangle$ = $13 \pm 1$\,mJy 
and $\langle$$ S_{9000}$$\rangle$ = $14.3 \pm 0.8$\,mJy, similar to the
flux density of $12 \pm 2$\,mJy obtained from the Methanol Multibeam pulsar survey
at a centre frequency of 6.3\,GHz. The spectrum therefore appears to have a positive spectral
index, highly unusual for pulsars which have a mean
spectral index of about --1.6 \citep{Lor95}. Again, this peculiar
spectral behaviour is observed in
the other two radio pulsating magnetars, which both
have a flat (or inverted) radio spectrum (\citealp{CamApJ679}; \citealp{Laz08}; \citealp{CamApJ669}).

\begin{table}[]
\caption{Parameter summary of PSR\,J1622-4950.}
\begin{tabular}{l l l}
\tableline
\tableline
&Parameter & Value\\
\tableline
\tableline
Observed &&\\
\tableline
&Right ascension (J2000) $^{\rm a}$		& 	16$^{\rm h}$22$^{\rm m}$44$^{\rm s}$.80(3)\\
&Declination (J2000) $^{\rm a}$			& 	--49$^{\circ}$50$'$54$''$.4(5)\\
&Galactic longitude $^{\rm a}$ 				& 	333.85\\
&Galactic latitude $^{\rm a}$ 				& 	--0.10\\
&Epoch									&	MJD 55080\\
&Spin period ($P$) 							& 	4.3261(1) s\\
&Period derivative ($\dot{P}$) 					& 	1.7(1)$\times$10$^{-11}$ s s$^{-1}$\\
&Dispersion measure ($DM$)				 	& 	820(30) cm$^{-3}$ pc\\
&Flux density at 1400 MHz ($S_{\rm 1400})$	  	&	4.8(3) mJy\\
&Rotation measure ($RM$)	 				&	--1484(1) rad m$^{-2}$\\
\tableline
Derived &&\\
\tableline
&Distance $^{\rm b}$						&	$\approx$ 9 kpc\\
&Surface magnetic field ($B$)				 	& 	2.8$\times$10$^{14}$ G\\
&Characteristic age	($\tau_{\rm c}$)				&	4 kyr\\
&Spin down luminosity ($\dot{E}$)			 	&	8.5$\times$10$^{33}$ erg s$^{-1}$\\
&X-ray luminosity ($L_{\rm X}$) $^{\rm c}$		&	2.5$\times$10$^{33}$ erg s$^{-1}$\\
\tableline
\tableline

\label{Tab:Summary}
\end{tabular}

$^{\rm a}$The position refers to the coordinates of the X-ray source.\\
$^{\rm b}$The distance is calculated from the dispersion measure with the Cordes--Lazio model.\\
$^{\rm c}$Assuming the dispersion measure derived distance.\\

\end{table}

\subsection{X-ray Emission}
For ordinary rotation-powered pulsars, the X-ray luminosity ($L_{\rm X}$)
is much smaller than the spin-down luminosity, $\dot{E} \equiv$ 4$\pi^2  I \dot{ P} P^{-3}$; 
on average $L_{\rm X} \approx 10^{-3}\dot{ E}$ \citep{Bec97}.
In contrast, magnetars are observed to have $L_{\rm X} \gtrsim \dot{E}$. 
An estimate of the unabsorbed 0.3-10\,keV flux for CXOU J162244.8--495054
gives 2.6 $\times$ 10$^{-13}$\,erg\,cm$^{-2}$\,s$^{-1}$, assuming a
Galactic hydrogen column density in that direction of $N_{\rm H}$ = 2 $\times$ 10$^{22}$\,cm$^{-2}$
\citep{Dic90} and a blackbody spectrum with $kT$ = 0.4\,keV, typical
of a quiescent magnetar. 
This implies $L_{\rm X}$(0.3-10\,keV) $\approx$ 2.5 $\times$ 10$^{33}$\,erg\,s$^{-1}$
at the estimated distance of 9\,kpc and, given
$\dot{ E} \approx$ 8.5 $\times$ 10$^{33}$\,erg\,s$^{-1}$ for this magnetar,
$L_{\rm X} \sim 0.3 \dot{E}$.
Since the distance calculated from the DM has a large uncertainty and is known only
to within a factor of two or so, the derived X-ray luminosity may be in error by up to a 
factor of four.
Even within these errors, the ratio is significantly higher than for ordinary pulsars, but well within
the range for the magnetars 1E\,1547--5408 and XTE\,J1810--197 which have
$L_{\rm X} \sim 0.1 \dot{E}$ and
$L_{\rm X} \sim 4 \dot{ E}$  (\citealp{Mer08}; \citealp{CamApJ666}; \citealp{Hal05}) in quiescence
and larger ratios in outburst.

The radio emission from XTE\,J1810--197 was discovered immediately following
a strong X-ray outburst. It has since faded, both in the radio and the X-ray band, and the
radio pulsations are no longer visible. For 1E\,1547--5408, the radio emission
is also highly variable and appears to be `revived' in the periods after
its X-ray outbursts. PSR J1622--4950 on the other hand, has had at least two episodes
of non-detections in the radio band lasting hundreds of days followed by periods
of bright radio emission (see Fig. \ref{Fig:Lightcurve}).
In addition to the new {\it Chandra} observation,
we searched archival data from \textit{Chandra}, \textit{XMM Newton}, \textit{Rosat}, \textit{ASCA}, \textit{Beppo-SAX},
{\it Rossi-XTE} and \textit{Swift} for an outburst, however no evidence for X-ray flux variability 
and no X-ray outburst at the level of the outbursts seen in XTE\,1810--197 and 1E\,1547--5408
in connection to the radio pulsations ($\gtrsim$ 10$^{36}$\,erg\,s$^{-1}$) were found since at least as early as 2005. 
It is possible therefore that an enhancement of X-ray activity is not a requirement for 
pulsed radio emission by magnetars, 
however, given the duty cycle of sensitive X-ray observations of the field containing PSR\,J1622--4950, 
we cannot constrain the occurrence of fainter X-ray enhancements of the source.
What is instead certain is that
the observed X-ray emission from PSR\,J1622--4950
is at variance with what is observed for the other two radio pulsating magnetars.

\subsection{Possible SNR Association}
If the true age of PSR\,J1622--4950 is similar to its characteristic age
of 4\,kyr, we might expect to see a supernova remnant (SNR) surrounding
the pulsar. Indeed, 5 of the 9 AXPs and at least 1 of the 5 SGRs are located within SNRs
(\citealp{Mer08}; \citealp{Gae01}).
Inspecting the ATCA image in Fig. \ref{Fig:ATCA-CXO}, we see a ring of emission centered $\sim$2$'$
 south of the pulsar location. This ring lacks an infra-red
counterpart and appears to be non thermal, whereas the extended radio 
source to the south of the ring is clearly thermal in nature.
Could the ring be the SNR and the pulsar has escaped its bounds?
If we assume a distance of $\sim$9\,kpc to the magnetar and further assume
it was born in the centre of the ring, the magnetar would need a velocity
of $\sim$1300\,km\,s$^{-1}$ to reach its current location whereas the ring
itself would have a lower expansion velocity. Such a velocity is high
(though not impossible) for pulsars but rather low for expanding SNRs.
Although the link between the ring and the magnetar is a possibility
we consider it unlikely.

\section{Conclusions}
The HTRU survey has discovered a radio-luminous pulsar, 
which is highly polarized, has an inverted spectrum, and is highly variable
in both its pulse profile and flux density.
The radio pulsar has a faint X-ray counterpart that appears to be stable in flux,
with a value that is typical of a quiescent magnetar. 
The pulsar shares many of
the properties of the two known radio magnetars and we therefore
conclude that PSR\,J1622--4950 is indeed a magnetar, the first discovered
through its radio emission.
This discovery not only adds a new
member to the magnetar family, 
but also highlights unprecedented features of the emission of the magnetars across the electromagnetic band. 
At odds with what is observed in other sources, PSR\,J1622--4950 indicates 
that bright radio emission can be present even when a magnetar displays an X-ray luminosity 
typical of a quiescent state. 
Moreover, PSR\,J1622--4950 shows that {\it pulsed} radio emission can either 
exist without the occurrence of a strong X-ray outburst, or occur a long time ($\gtrsim$ 5 years) after the outburst. 
Alternatively, the radio pulsations could be triggered by a modest increment of X-ray activity, that escaped detection in this case.
We finally note that the extreme variability in the flux density of PSR\,J1622--4950 also
demonstrates the advantages of surveying the radio sky
at regular intervals with even modest sensitivity. 
This highlights the potential of the upcoming radio facilities like the LOFAR, 
ASKAP or the SKA which promise to characterize the dynamic radio sky at an unprecedented level.

\acknowledgments
The Parkes Observatory and the Australia Telescope Compact Array are part of the Australia Telescope, which is funded by the Commonwealth of Australia for operation as a National Facility managed by CSIRO. The Chandra X-ray Observatory Centre is operated by the Smithsonian Astrophysical Observatory for and on behalf of the National Aeronautics Space Administration under contract NAS8-03060. This work is partly supported by the Australian Research Council through its discovery programme. 
The HYDRA supercomputer at the 
JBCA is supported by a grant from the UK Science and Technology Facilities Council. S.B. 
gratefully acknowledges the support of STFC in his PhD studentship. This work is partly 
supported by the Australian Research Council through its discovery programme.

{\it Facilities:} \facility{Parkes}, \facility{ATCA}, \facility{CXO (ASIS-I)}.

\end{document}